\documentstyle[12pt,aaspp4,flushrt,epsf]{aastex}

\newcommand{\be}	{\begin{equation}}
\newcommand{\ee}	{\end{equation}}
\newcommand{\beq}	{\begin{eqnarray}}
\newcommand{\eeq}	{\end{eqnarray}}
\newcommand{\bd}	{\begin{displaymath}}
\newcommand{\ed}	{\end{displaymath}}
\newcommand{\define}	{\stackrel{\rm def}{=}}

\begin{document}

\title{\bf Stochastic Coagulation and the Timescale for Runaway Growth}

\author{Leonid~Malyshkin and Jeremy~Goodman}
\affil{	Princeton University Observatory, Princeton, NJ~08544 \\
	E-mail:~leonmal@astro.princeton.edu}
\date{}

\begin{abstract}
We study the stochastic coagulation equation using simplified models
and efficient Monte Carlo simulations.  It is known that (i) runaway
growth occurs if the two-body coalescence kernel rises faster than
linearly in the mass of the heavier particle; and (ii) for such
kernels, runaway is instantaneous in the limit that the number of
particles tends to infinity at fixed collision time per particle.
Superlinear kernels arise in astrophysical systems where gravitational
focusing is important, such as the coalescence of planetesimals to
form planets or of stars to form supermassive black holes.  We find
that the time required for runaway decreases as a power of the
logarithm of the the initial number of particles.  Astrophysical
implications are briefly discussed.

{\it Key Words}: planetesimals, planetary formation; collisional physics; 
methods, numerical.
\end{abstract}


\section{Introduction}\label{INTRODUCTION}

A frequently-encountered process in many fields of
science is the random coalescence of small bodies into larger ones,
conserving total mass.  Astrophysical examples include
the coalescence of planetesimals into planets \citep{Safronov}
and of stars into black holes \citep{Lee,QS}.
When the number of bodies is large, coalescence is often modeled
by Smoluchowski's equation \citep{Smo}, also known as the
\emph{statistical equation}:
\be\label{eq:Smo}
\frac{dc_i}{d\tau} = \frac{1}{2}\sum\limits_{j=1}^{i-1} K(i-j,j)c_{i-j}c_j
-c_i\sum\limits_{j=1}^{\infty} K(i,j) c_j,
\ee
Here the \emph{concentration} $c_i(\tau)$ is the number of bodies
per unit volume of mass $m_i\propto i$ at time $\tau$, and the functional
form of the \emph{coagulation kernel} $K(i,j)=K(j,i)$ is chosen
to approximate the mass dependence of the two-body collision rate.

The statistical equation
is intended to model coalescence on average, smoothing over
fluctuations.  It is not expected to be accurate when the initial
number of particles, $N$, is small.
More interestingly, eq.~(\ref{eq:Smo}) can fail even for 
$N\gg1$ if $K(i,j)$ increases sufficiently rapidly with $i$ and $j$.
Coalescence should conserve mass, so that the mass density
\bd \rho(\tau)\define
\sum\limits_{k=1}^\infty k\, c_k(\tau)
\ed
is constant.
It is easy to show from eq.~(\ref{eq:Smo}) that
$\dot\rho=0$ provided that the relevant summations converge and can be
interchanged.
But in the analytic solution of the case $K(i,j)=ij$ starting from
the \emph{monodisperse} initial conditions
\be \label{monodisperse}
c_i(0)=\delta_{i1},\qquad\rho(0)=\frac{N}{V}=1,
\ee
$\rho(\tau)$ begins to decrease  after $\tau=1$ \citep{Trubnikov}.
This is usually interpreted to mean that a macroscopic \emph{runaway
particle} has formed---also known as a \emph{gel}, because of
applications in physical chemistry.
It is believed that for $K(i,j)\sim (ij)^\nu$ with $\nu>1$, gelation
begins immediately in the statistical equation
\citep[and references therein]{Jeon,MHL}.
The model~(\ref{eq:Smo}) can be extended
to include the gel/runaway particle explicitly \citep{Flory}.
Monte Carlo simulations for large but finite $N$
show, however, that such models do not accurately predict the time
dependence of the gel mass even on average if $\nu>1$ \citep{Spouge}.

Coagulation is more correctly described by the
\emph{stochastic} equation for the joint probability $f(n_1,n_2,\ldots;t)$
for the occupation numbers $\{n_1,n_2,\ldots\}$ of the mass bins:
\beq
\frac{\partial}{\partial t} f({\bf n}; t)&=&\sum\limits_{i=1}^N \,
\sum\limits_{j=i+1}^N K(i,j)\,(n_i+1)(n_j+1)
 f(...,n_i+1,...,n_j+1,...,n_{i+j}-1,...; t)\nonumber\\
&+&\sum\limits_{i=1}^N \frac{1}{2} K(i,i)\,(n_i+2)(n_i+1)
 f(...,n_i+2,...,n_{2i}-1,...; t)\nonumber\\
&-&\sum\limits_{i=1}^N \,\sum\limits_{j=i+1}^N K(i,j)\,n_i n_j
f({\bf n}; t)~-~
\sum\limits_{i=1}^N \frac{1}{2} K(i,i)\,n_i(n_i-1) f({\bf n}; t),
\label{STOCHASTIC_EQ}
\eeq
The statistical equation results
from taking first moments,
\be\label{avgn}
c_i(t)\define V^{-1}\,\overline{n}_i(t)\define 
V^{-1}\,\int n_i f(n_1,\ldots,n_{i-1},n_{i},n_{i+1},\ldots; t)
\,dn_1\ldots dn_{i-1} dn_{i+1}\ldots,
\ee
in the limit that $N,V\to\infty$ at fixed $\rho(0)$.
In order to get a closed set of equations, one assumes that
$\overline{n_i n_j}=\overline{n}_i\times\overline{n}_j$.  This is justified
if the occupation numbers are approximately uncorrelated.
But if runaway should occur, then the occupation numbers of all
the low-mass bins are correlated with bins $i\sim N$ traversed
by the runaway particle.

Like the statistical equation, the stochastic equation has analytic
solutions for the kernels
$K(i,j)\propto~\mbox{constant}$, $i+j$, and $i\times j$
\citep{Lushnikov, TN}.
For the first two kernels, the predictions of the two equations
are compatible in the following sense:
after monodisperse initial conditions, $c_i(\tau)$ computed from
eq.~(\ref{eq:Smo}) agrees with $c_i(t)$ computed from
eqs.~(\ref{STOCHASTIC_EQ})-(\ref{avgn}) in the limit $N\to\infty$
if $t$ and $\tau$ are related by the initial collision time
per particle,
\be
t_{\rm coll}\define\left[(N-1)\,K(1,1)\right]^{-1},\qquad
\tau= t/t_{\rm coll}. \qquad 
\label{DIMENSIONAL_TIME}
\ee
For these kernels, there is no runaway, and the statistical equation
conserves mass.
Even for $K(i,j)=ij$, a similar agreement is found between
the statistical and stochastic equations at $\tau<1$
and mass bins $i\ll\sqrt{N}$ \citep{TN}.  The stochastic
equation confirms that runaway begins at $\tau>1$;
more precisely, \cite{Lushnikov} shows that the quantity
\bd
\kappa(\tau)\define \lim_{N\to\infty}\left[N^{-2}
\sum\limits_{i=1}^N\, i^2\,\overline{n}_i(\tau)\right],
\ed
which can be interpreted as the runaway mass fraction on average,
becomes nonzero at $\tau=1$ and $\approx 1-2e^{-\tau}$
at $\tau\gg 1$.

Kernels such that $K(i,j)\propto i^\nu$ when $i\gg j$ are sometimes
called ``unphysical'' if $\nu>1$ because if collision rate were
proportional to surface area, then surface area would have to increase
more rapidly than mass.  But in astrophysics (and elsewhere),
gravitational focusing and mass stratification may conspire to produce
$\nu>1$ (\S5).  Thus a better term for such kernels would be
\emph{superlinear}.  The occurrence and timescale of runaway are of
great interest since the runaway particle may represent a planet or
supermassive black hole, for example.  Unfortunately, few general
results are known for the superlinear regime.  \cite{Spouge}
conjectured on the basis of Monte Carlo simulations of stochastic
coagulation that {\it runaway is instantaneous for $\nu>1$ in the
limit $N\to\infty$}: in other words, the entire mass is consumed by a
single particle after an infinitesimal multiple of the single-particle
collision time $t_{\rm coll}$.  A similar conjecture had previously
been made by \cite{DLP} for the special case $\nu=3$, also on the
basis of Monte-Carlo simulations.  Recently, \cite{Jeon} has supplied
a proof.

Clearly, runaway/gelation is not instantaneous for finite $N$.
To the best of our knowledge, there have been no general and quantitative
statements about the scaling of the runaway time with $N$, although
the question is an important one since
$N$ is never truly infinite in practical applications.
The main result of the present paper will be
the conjecture, supported by Monte Carlo simulations, that
$\tau_{\rm runaway}$ varies as a negative power of $\log N$
for $\nu>1$.  Even before
performing our simulations, we were lead to this conjecture on the basis
of the highly simplified model presented in \S 2.  Our Monte Carlo algorithm
for stochastic coagulation is described in \S3 and tested against the 
analytic solutions of the stochastic equation cited above.  Simulations
for $\nu>1$ are reported in \S4.  Finally, in \S5, we discuss applications
of our conjecture to cases of astrophysical importance.


\section{MONOTROPHIC MODEL}\label{TOY_MODEL}

The full stochastic equation seems to be analytically
intractable for general values of the merging exponent
$\nu$.
In this section we solve a simplified problem
in which a single ``predator'' feeds upon
an unevolving population of unit-mass ``prey.''
The prey do not merge with one another but only with the
predator, whose initial mass is equal to that of the prey.
A slight additional simplification results from assuming that the
prey population is infinite.
We call this model ``monotrophic'' after the Greek
\emph{mono} (one) $+$ \emph{trophein} (to nourish).
Our Monte-Carlo simulations of the full stochastic equation
are well described by the monotrophic model in their later stages
when (or if) the most massive particle exhibits runaway growth.

We work in time units $\tau$ such that the feeding rate
of a unit-mass predator is unity, and 
we assume that the feeding rate increases as the $\nu^{\rm th}$ power
of its mass.
Let the predator start with mass equal one at time $\tau=0$, and 
let $p(k,\tau)$ be the probability that it is in the $k^{\rm th}$ mass bin 
and correspondingly has mass $m_k=k$ after time $\tau$.
The probability that the predator will enter mass bin $k$
during the brief time interval $(\tau,\tau+d\tau)$
is clearly $p(k-1,\tau)\times (k-1)^\nu\times  d\tau$, and the probability
that it will vacate bin $k$ during the same interval is 
$p(k,\tau)\times k^\nu\times d\tau$.
Hence the evolutionary equation and initial conditions for $p$ are
\be\label{P_EQN}
\frac{dp}{d\tau}(k,\tau) = (k-1)^\nu p(k-1,\tau) - k^\nu p(k,\tau),
\qquad\mbox{and}\quad p(k,0) = \delta_{k,1}.
\ee
One might suppose that
$\displaystyle \frac{d}{d\tau}\sum\limits_{k=1}^\infty p(k,\tau)=0, $
since the sum of the righthand sides (\ref{P_EQN}) would appear to cancel,
but after solving eqs.~(\ref{P_EQN}), one finds that
\be\label{PHI_DEF}
\phi(\tau)\define 1-\sum\limits_{k=1}^\infty p(k,\tau)
\ee
can be nonzero for $\tau>0$.
Clearly $\phi(\tau)$ is the probability that the predator is
not to be found in any finite mass bin.
So we interpret $\phi(\tau)$ as the probability that runaway growth has
occurred.
From eq.~(\ref{P_EQN}), $\phi(0)=0$ and
\be\label{LIMIT}
\frac{d\phi}{d\tau}(\tau)= \lim_{k\to\infty} k^\nu p(k,\tau).
\ee
Hence, the runaway cannot occur until the mass probability
function develops a power-law tail $p(k,\tau)\propto k^{-\nu}$.
But $\sum_{k=1}^\infty p(k,\tau)$ must be finite (in fact $\le 1$),
and the power-law tail has a convergent sum only if
$\nu>1$.  So we conclude that runaway occurs in the monotrophic
model only if $\nu>1$.

To solve for $\phi(\tau)$, we take the
Laplace transform $\tau\to z$ of eq.~(\ref{P_EQN}),
\bd
z\tilde p(k,z)~-\delta_{k,1} = (k-1)^\nu\tilde p(k-1,z) -k^\nu\tilde p(k,z).
\ed
This is easily solved:
\be\label{PT_SOL}
k^\nu \tilde p(k,z) = \prod_{\ell=1}^k \left(1+\frac{z}{\ell^\nu}\right)^{-1}
~~~~(k\ge 1).
\ee
In view of eq.~(\ref{LIMIT}), the Laplace transform of $\phi$ is
\be\label{PHIT_SOL}
\tilde\phi_\nu(z)= \frac{1}{z}\;\lim_{k\to\infty}\;\prod_{\ell=1}^k
\left(1+\frac{z}{\ell^\nu}\right)^{-1}.
\ee
A subscript has been placed on $\tilde\phi_\nu(z)$ to acknowledge its
dependence on $\nu$.
Standard convergence tests show that the limit of the product vanishes
if $\nu\le 1$ except at the poles $z=-1^\nu,-2^\nu,-3^\nu,\ldots$,
so that $\phi_{\nu\le 1}(\tau)=0$ also vanishes.
Once again, therefore, we see that
runaway does not occur  for $\nu\le 1$.
On the other hand, if $\nu>1$ then the product has a finite
nonzero limit.  

Exact results are possible for $\nu=2$.  The product~(\ref{PHIT_SOL}) has
the closed form
\be\label{PHIT_2}
\tilde\phi_2(z)= \frac{\pi}{\sqrt{z}\sinh(\pi\sqrt{z})}.
\ee
The only singularities of this function at finite $z$ are simple
poles on the negative real axis where $\sqrt{-z}$ is a positive integer.
Evaluating the inverse Laplace transform by residues yields an infinite
series that converges rapidly at large $\tau$:\footnote{
$\phi_2(\tau)$ can also be 
expressed in a parametric closed form involving elliptic integrals.}
\be\label{PHI_2}
\phi_2(\tau)= 1+2\sum\limits_{k=1}^\infty (-1)^k e^{-k^2 \tau}
\ee
Writing
\begin{eqnarray*}
\phi_2(\tau)&=&\int\limits_{-\infty}^\infty \Delta(x)\,e^{-\tau x^2}dx,\\
\mbox{where}~~~
\Delta(x)&\define&\sum\limits_{k=-\infty}^\infty (-1)^k \delta(x-k)
   ~~=~~ 2 \sum\limits_{n=0}^\infty\cos\left[(2n+1)\pi x\right],
\end{eqnarray*}
and integrating the Fourier series term by term yields a series
that converges rapidly at small times:
\be\label{PHI_2p}
\phi_2(\tau)=2\sqrt{\frac{\pi}{\tau}}\,\sum\limits_{n=0}^\infty
\exp\left[-\frac{\pi^2(2n+1)^2}{4\tau}\right].
\ee
Eq.~(\ref{PHI_2p}) shows
that runaway is exponentially unlikely for $\tau\ll 1$, while
eq.~(\ref{PHI_2}) shows that it is virtually
certain for $\tau\gg 1$. The probability reaches $50\%$
at $\tau\approx 1.37$.

Similar statements hold for general values of $\nu>1$ (cf. the Appendix):
\beq\label{PHINU_APPROX}
\phi_\nu(\tau)\approx\cases{
 (2\pi)^{(\nu-1)/2}\sqrt{\nu/(\nu-1)\tau}\,
\exp\left\{-(\nu-1)\left[\nu\sin(\pi/\nu)/\pi\right]^{-\nu/(\nu-1)}
\tau^{-1/(\nu-1)}\right\} & if $\tau\ll 1$,\cr
~ & ~ \cr
1~-~\Big[\prod_{k=2}^\infty\left(1-k^{-\nu}\right)\Big]^{-1}\,
\exp(-\tau) & if $\tau\gg 1$.}
\eeq
These expressions agree with the leading
terms of the series (\ref{PHI_2}) and (\ref{PHI_2p}) when $\nu=2$.
Scrutiny of the first exponential in eq.~(\ref{PHINU_APPROX}) reveals
that as $\nu$ approaches unity from above, the time at
which the runaway probability first becomes appreciably different from zero
is $\approx(\nu-1)^{-1}$.

In the original model of \S 1, all particles can be  predator or
prey.  For $\tau\ll 1$, the results of this section suggest that
the probability that any given particle has achieved large mass
is $\approx\phi(\tau)$.
We presume these probabilities to be approximately independent if $N$ is
sufficiently large, because
two randomly selected particles could both grow to a mass $\gg 1$
before competing for the same prey.
Thus the \emph{total} probability of runaway 
at a time $\tau<<1$ is $\approx N\phi(\tau)$.
Given the  estimate (\ref{PHINU_APPROX}) for $\phi(\tau)$,
the total runaway probability approaches unity at a time scaling as
\be
\tau_{\rm runaway}\propto {(\log N)}^{\gamma(\nu)}
\label{ESTIMATED_RUNAWAY_TIME}
\ee
with $\gamma(\nu)=1-\nu<0$, but only if $\nu>1$.  
To test this conjecture, we undertook the Monte Carlo simulations described
below.


\goodbreak
\section{Numerical Technique and Tests}\label{TECHNIQUE}

Direct numerical solutions of the stochastic coagulation 
equation~(\ref{STOCHASTIC_EQ}) are difficult to obtain for large $N$,
so we have used Monte Carlo simulations.
Great savings in memory and processor time can be achieved by keeping
track of the occupied bins only.  Therefore, we adopt an
indexing scheme different from that of \S 1:  at each time step, $n_i\ne0$ 
will represent the occupation number of the $i^{\rm th}$ \emph{nonempty}
bin, with mass per particle $m_i$, and $1\le i\le I$.
For expository convenience, the bins are sorted by mass 
($m_i<m_j$ if $i<j$), although this is not essential to the algorithm.
The total mass is
\be
\sum\limits_{i=1}^I\,n_i\,m_i=N.
\label{TOTAL_MASS}
\ee
The storage required by our scheme is $O(I)$.
Because $m_i\ge i$ and $n_i\ge 1$, the summation (\ref{TOTAL_MASS})
is $\ge I(I+1)/2$.  Hence the number of occupied bins $I<{(2N)}^{1/2}$.
In our simulations, the maximum $I$ encountered is
almost always much smaller than this upper bound.

The coagulation kernel is now written as
$K(m_i,m_j)$ rather than $K(i,j)$ to emphasize that the coalescence
rate depends upon the particle masses rather than the arbitrary bin indices.
The total coalescence rate of bins $i,j\ge i$ is
\be\label{IJRATE}
R_{i,j}\define K(m_i,m_j)\times\cases{n_i n_j &  ~$i<j$\cr
   			              n_i(n_i-1)/2 & ~$i=j$}.
\ee
It is useful to imagine these rates arranged as an 
$I\times I$ upper-triangular matrix, and to introduce the
partial sums
\be\label{SUMIJ}
S_{i,j} \,\define~ \sum\limits_{k=1}^{i-1}\sum\limits_{\ell=k}^I\,R_{k,\ell} + 
		   \sum\limits_{\ell=i}^j\,R_{i,\ell},
\ee
\emph{i.e.} $S_{i,j}$ is the sum of the first $i-1$ rows plus the
first $j$ columns of row $i$,
and the total coalescence rate $S\define S_{I,I}$.
We store in the computer memory the rates $S_{i,I}$, $i=1,2,...I$.
Each merging alters the occupation numbers of at most
three bins, so these rates can be updated in $O(I)$ operations.
Indirect indexing minimizes the work of inserting or deleting
rows and columns as the list of occupied bins changes.

Every simulation begins with monodisperse initial conditions,
\be
n_1=N,\quad
m_1=1,\quad
I=1,
\label{INITIAL_CONDITIONS}
\ee
and ends after $N-1$ mergings with
\be
n_1=1,\quad
m_1=N,\quad
I=1.
\label{FINAL_CONDITIONS}
\ee
Merging occurs at random
but increasing times $t_1<t_2\ldots\le t_{N-1}$ chosen as follows:
If ${S}$ is the total coalescence rate computed
after the $s^{\rm th}$ merging, the probability that no further
merging occurs before $t>t_s$ is
\bd
{\cal P}=\exp\left[-{S}\cdot(t-t_s)\right].
\ed
Therefore, we chose a random number $X\sim U(0,1]$ (\emph{i.e.},
$X$ is uniformly distributed between $0$ and $1$) and take
\bd
t_{s+1}=t_s~-{S}^{-1}\ln{X}.
\ed
Thus merging occurs at $t_{s+1}$, and the next task is to decide
which bins are involved.
Choosing a second random number $Y\sim U(0,1]$, we find $i,j$ 
($j\ge i$) such that
\be
S_{i,j-1}~<~Y\cdot S~\le~S_{i,j}.
\ee
Since $S_{i,j}-S_{i,j-1}=R_{i,j}$ [eq.~(\ref{SUMIJ})],
bins $i,j$ are selected with the correct probability
$R_{i,j}/S$,
all rates being evaluated just before the $s+1^{\rm st}$ merging.

With this scheme, there are $N-1$ steps per simulation and
$O(I)$ operations per step, so the number of
operations per simulation is at most $O(N^{3/2})$.  In practice,
$I\ll N^{1/2}$ so that the computer time
is almost linearly proportional to $N$.
We average many simulations to obtain the statistics of
quantities of interest.

We have tested our code against the exact solutions
of \cite{TN} for the cases
$K(m_i,m_j)=1$ and $K(m_i,m_j)=m_i+m_j$.
We compare at $N=10$ and find excellent agreement
(Fig.~{\ref{FIGURE_COMPARISON}).  Not surprisingly, even though
runaway does not occur, the statistical model is a poor approximation
at such small $N$.

\begin{figure}
\vspace{6.0cm} \includegraphics{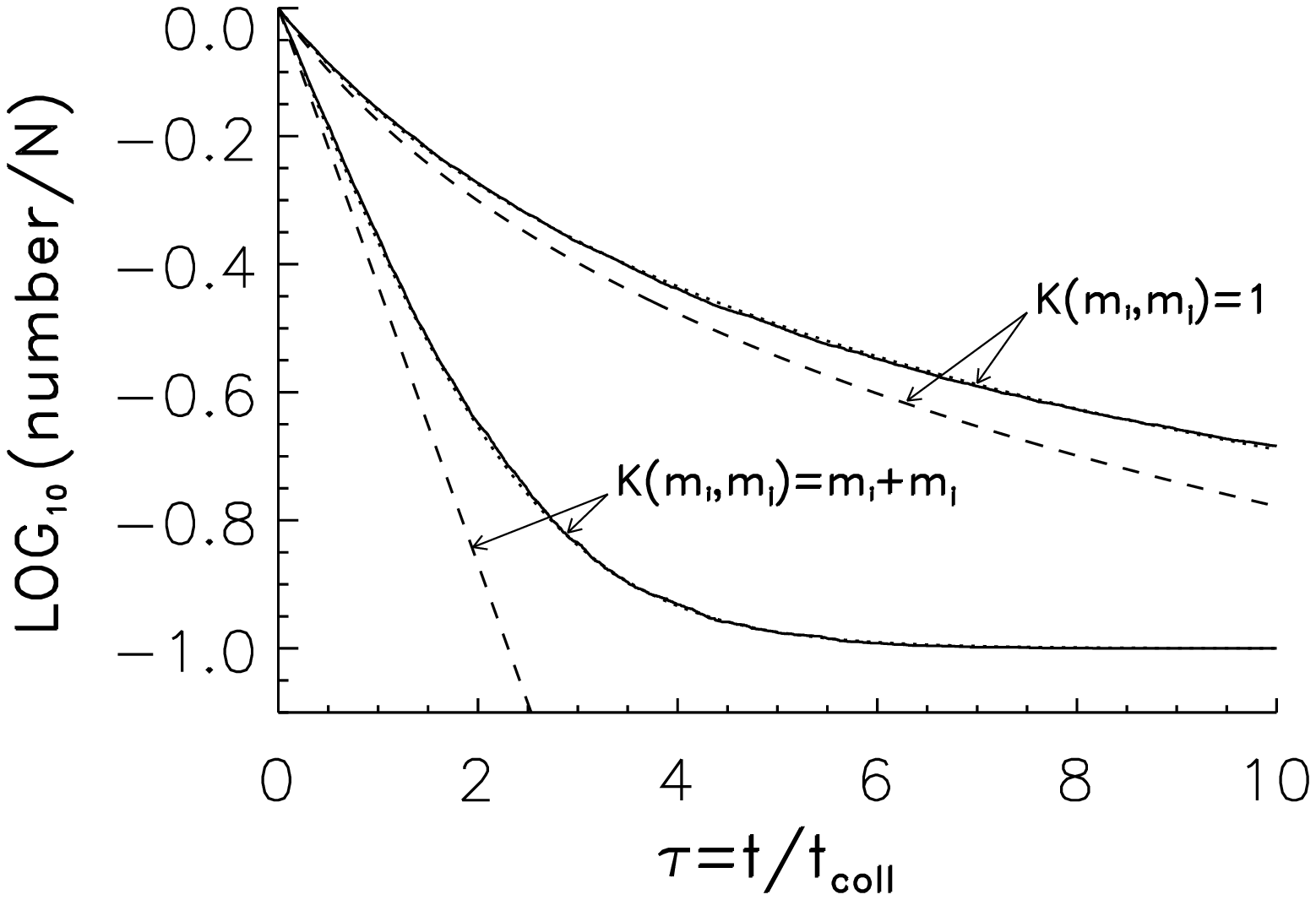}
\caption{ The number of particles, averaged over $1000$ simulation runs
and normalized to the initial number of particles ($N=10$), versus our
dimensional time is shown by the two solid lines. The analytical
results for the stochastic and statistical coagulation equations are
shown by the two dotted lines (which almost coincide with the
corresponding solid lines) and two dashed lines respectively.}
\label{FIGURE_COMPARISON}
\end{figure}


\section{SIMULATION RESULTS}\label{RESULTS}

We discuss simulations for
two classes of coalescence rates, $K(m_i,m_j)={(m_i\times m_j)}^\nu$ 
(\emph{multiplicative} kernel) and $K(m_i,m_j)={(m_i+m_j)}^\nu$, $\nu>1$
(\emph{additive} kernel).

\begin{figure}
\vspace{18.15cm}
\includegraphics{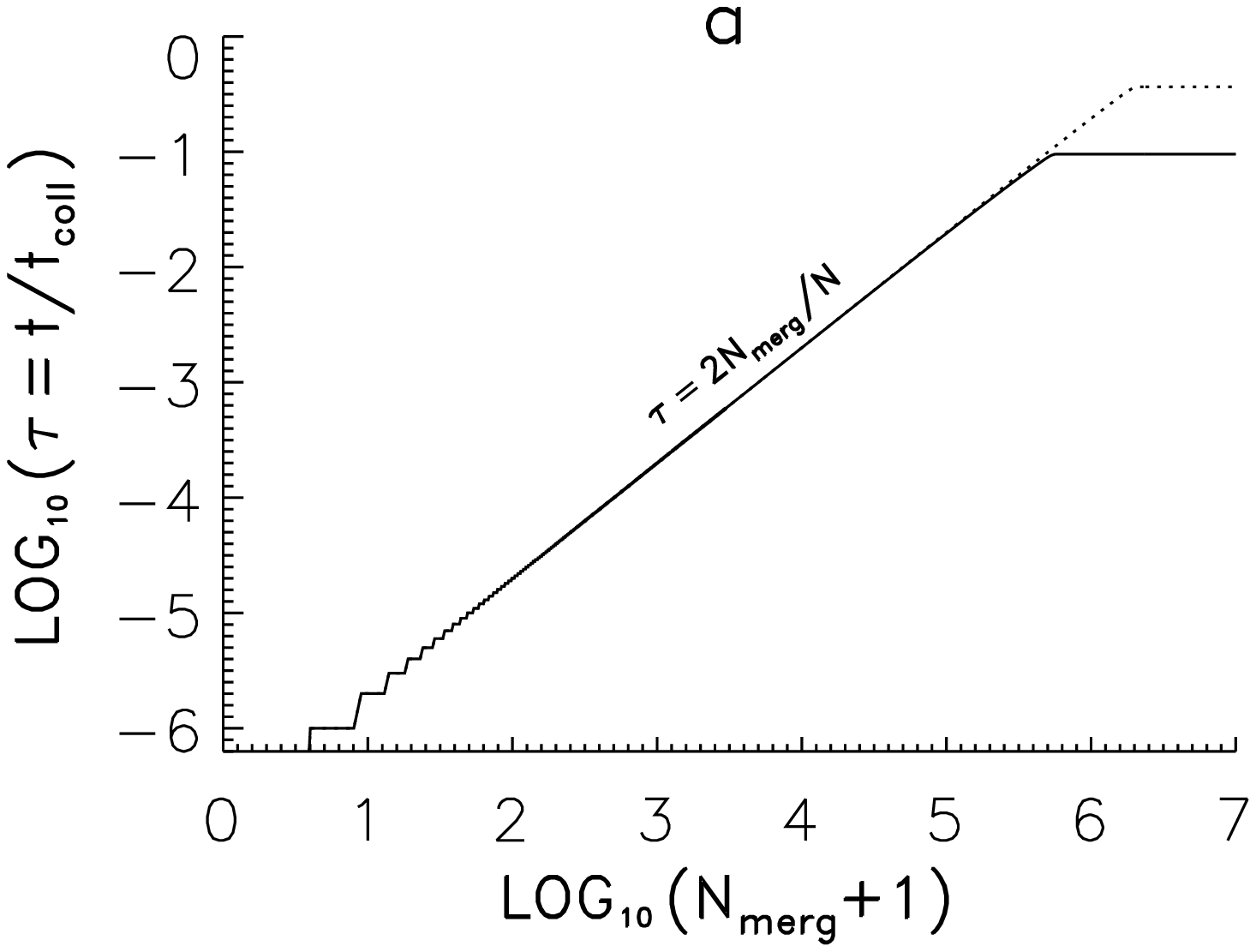}
\includegraphics{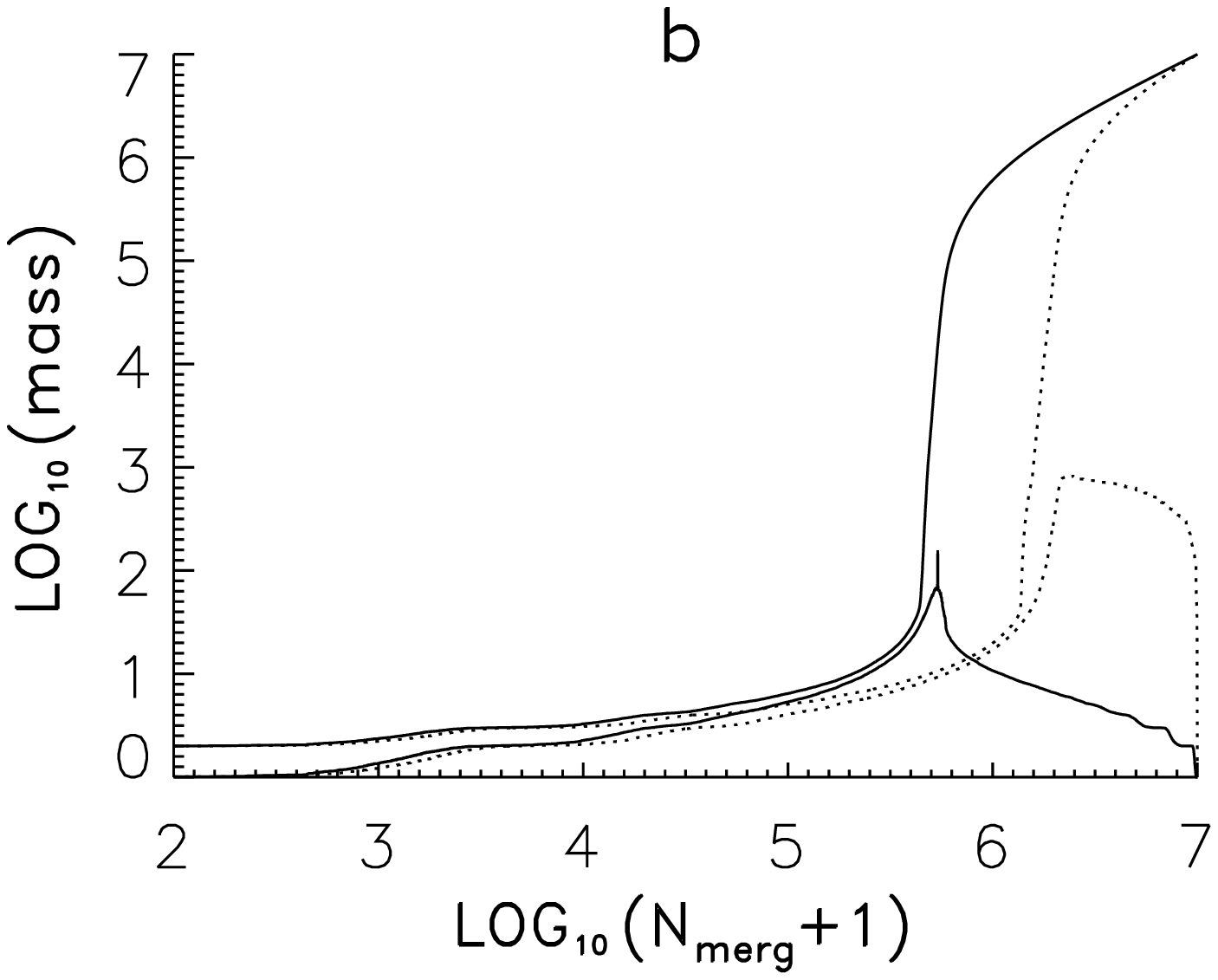}
\includegraphics{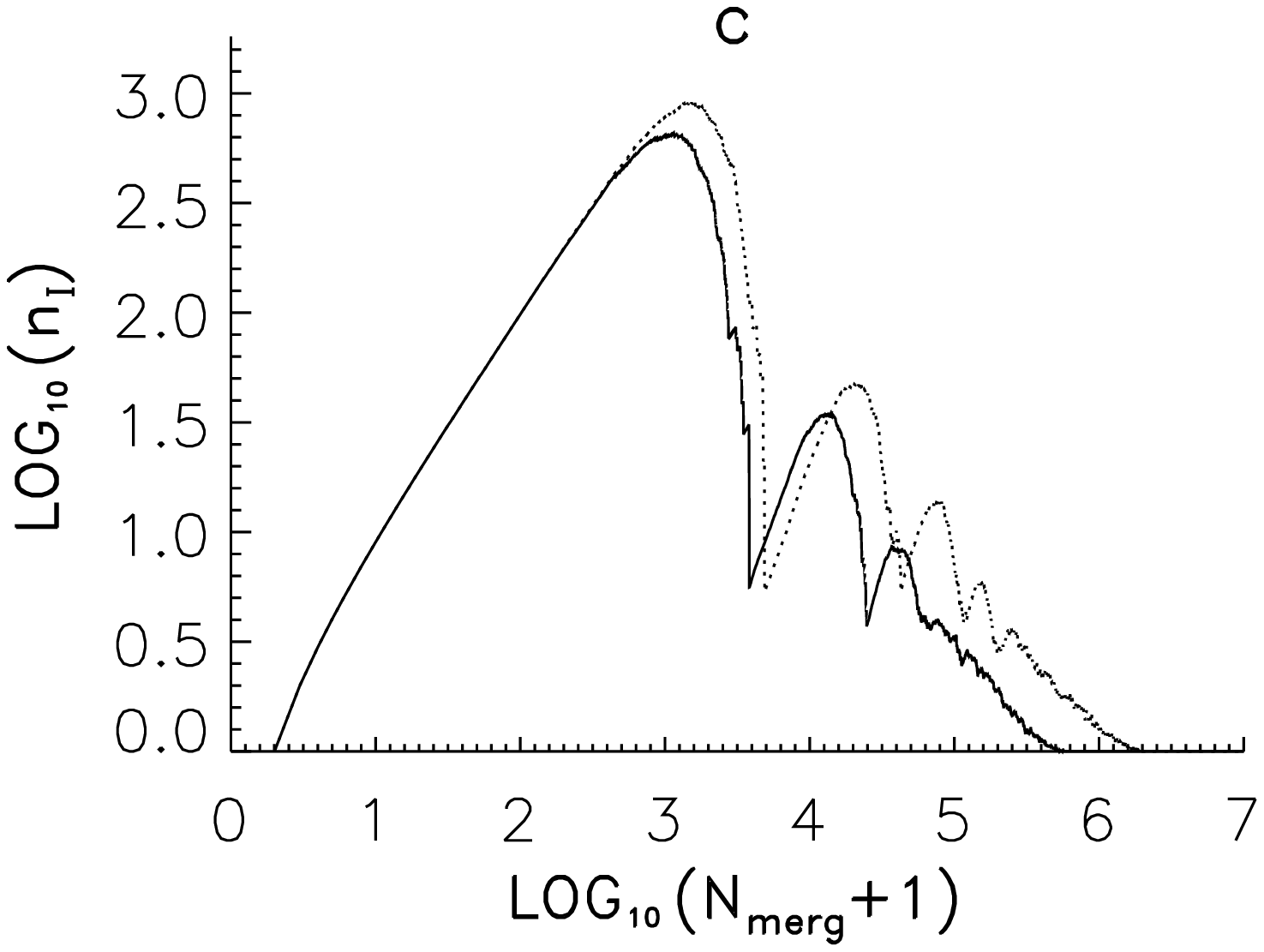}
\includegraphics{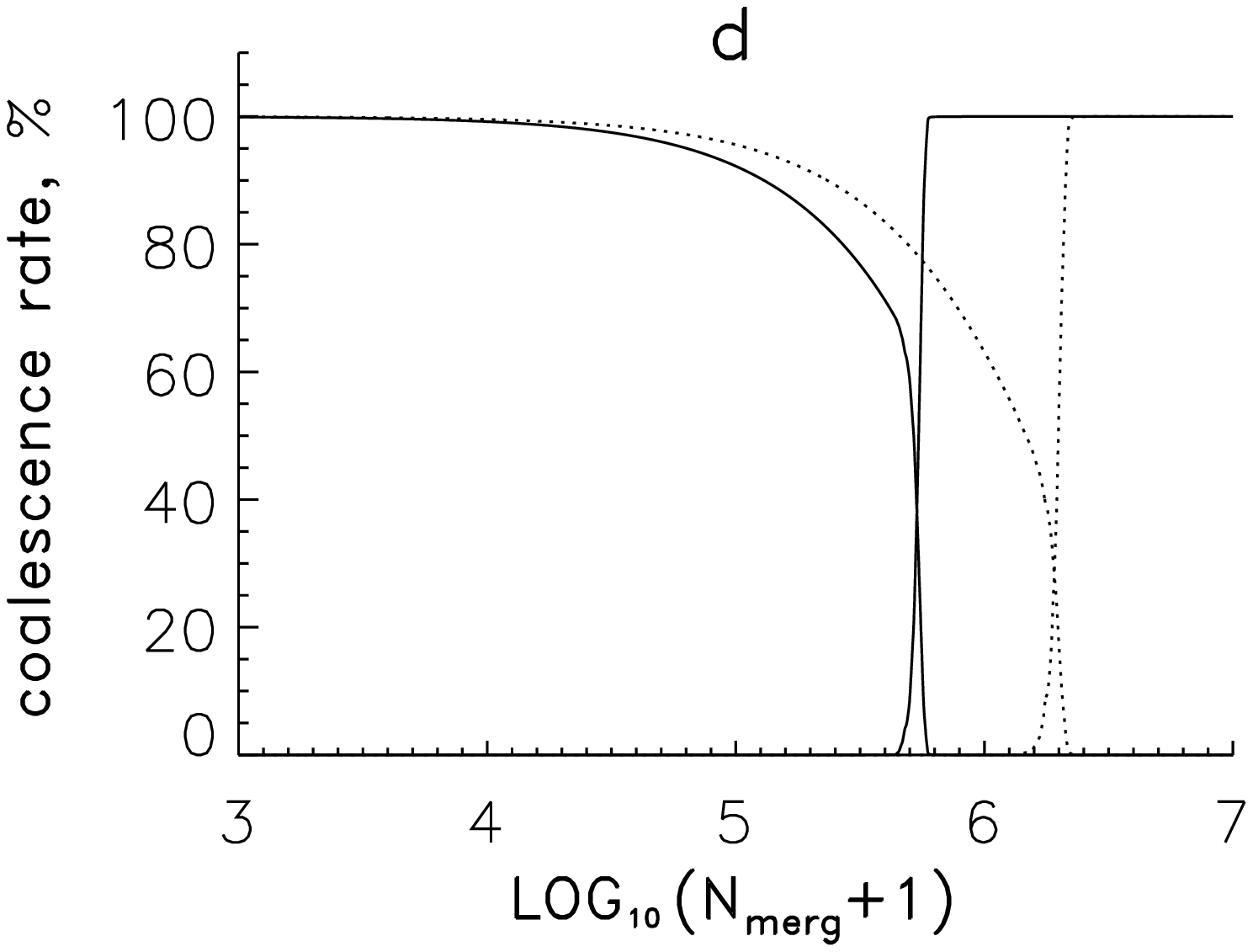}
\includegraphics{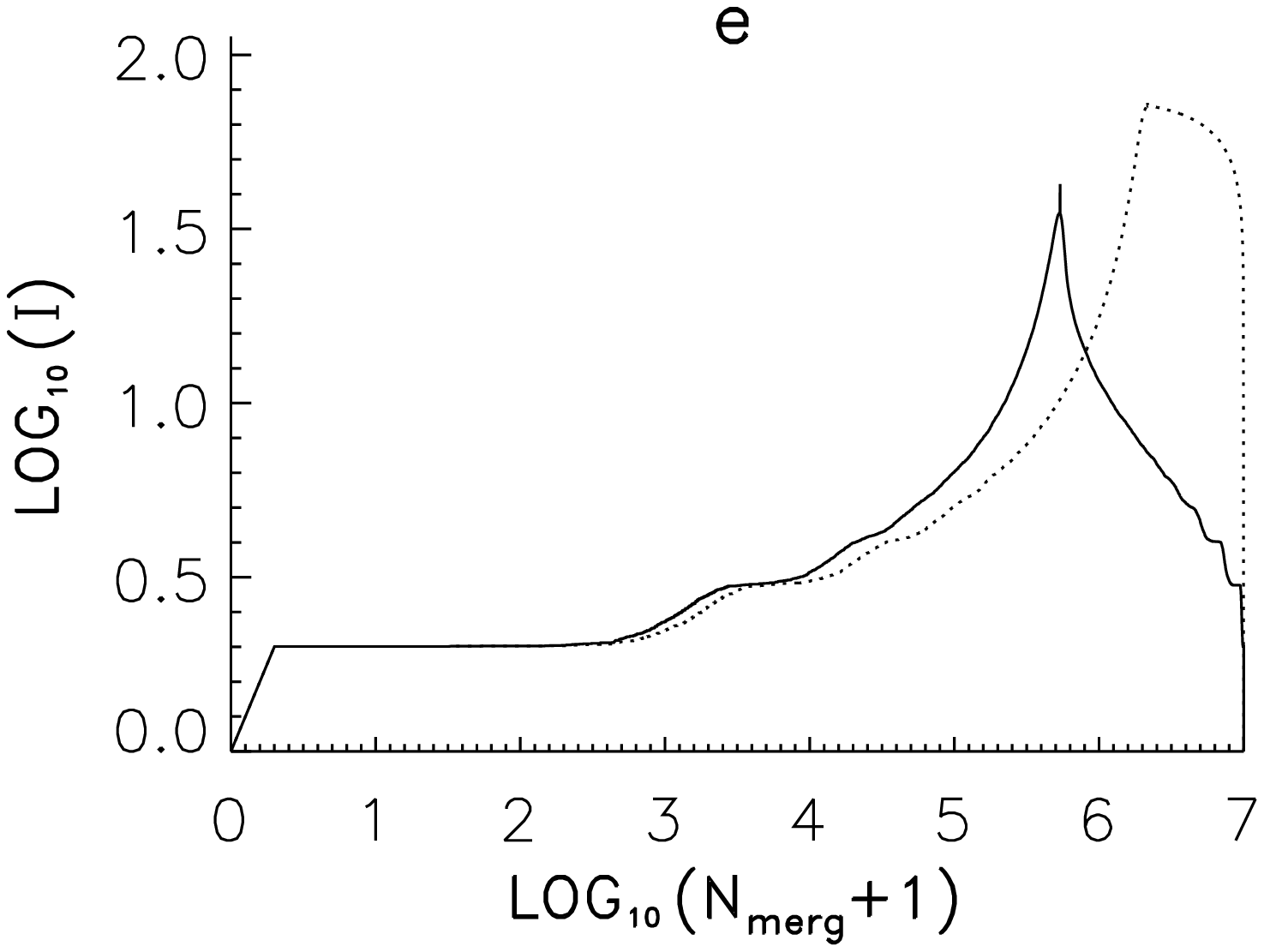}
\includegraphics{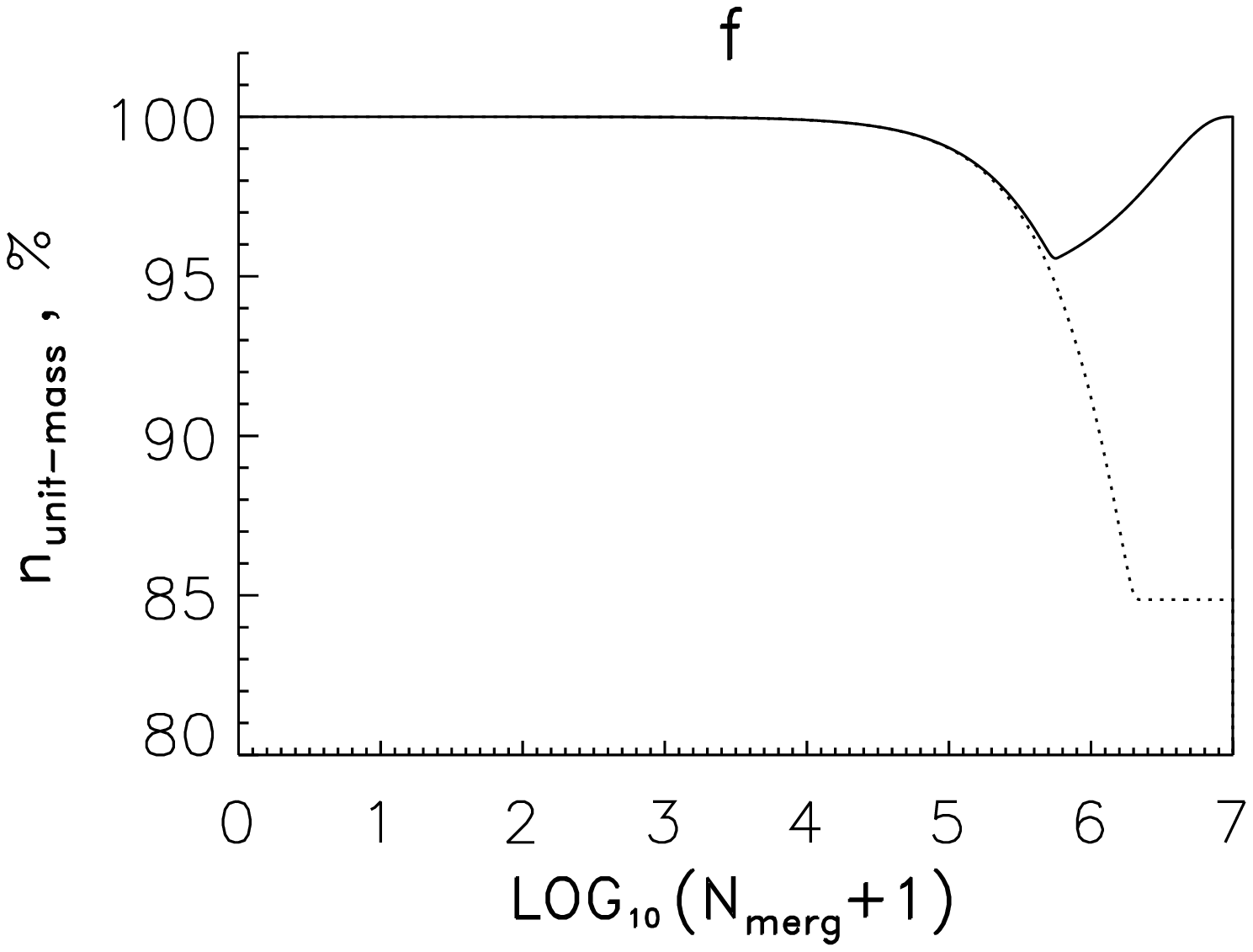}
\caption{
Average evolution for $300$ simulations with $N=10^7$.
Solid lines: $K(m_i,m_j)=(m_i\times m_j)^2$; dotted lines:
$K(m_i,m_j)=(m_i+m_j)^2$.
(a): time versus number of mergings, $N_{\rm merg}$;
(b):~the mass per particle of the highest (monotonic curves) and 
the second highest bins, \emph{i.e.} $m_I$ and $m_{I-1}$;
(c):~the number of particles in the highest bin;
(d):~fraction of the total coalescence rate that involves one 
particle taken from the highest bin (monotonically increasing),
and by mergings between unit-mass particles (decreasing);
(e):~number of occupied mass bins, $I$;
(f):~number of unit-mass particles as a fraction of the total
remaining, \emph{i.e.} $\delta_{1,m_1}\,n_1/(N-N_{\rm merge})$.}
\label{RUNAWAY_PARAMETERS}
\end{figure}

\begin{figure}
\vspace{6.6cm}
\includegraphics{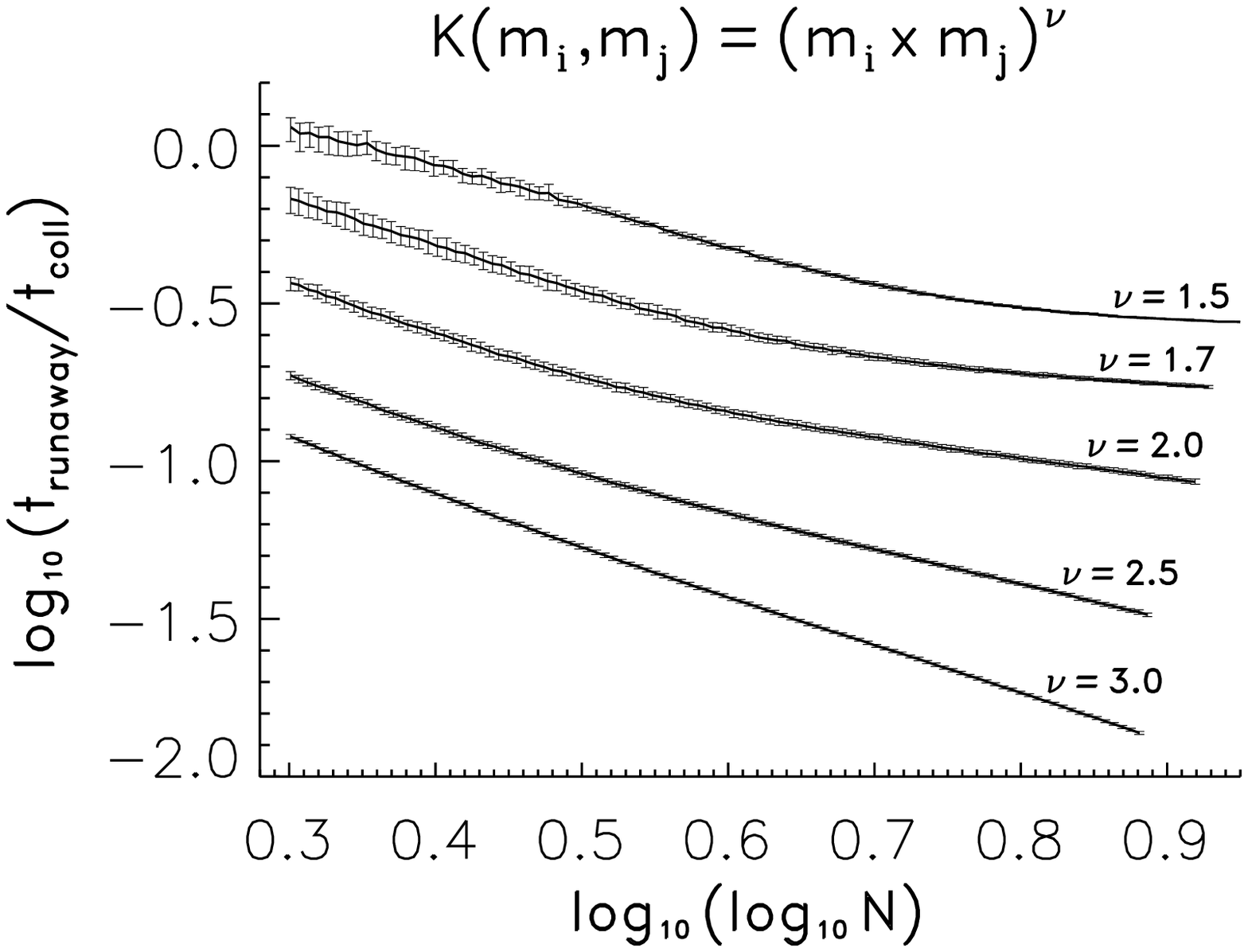}
\includegraphics{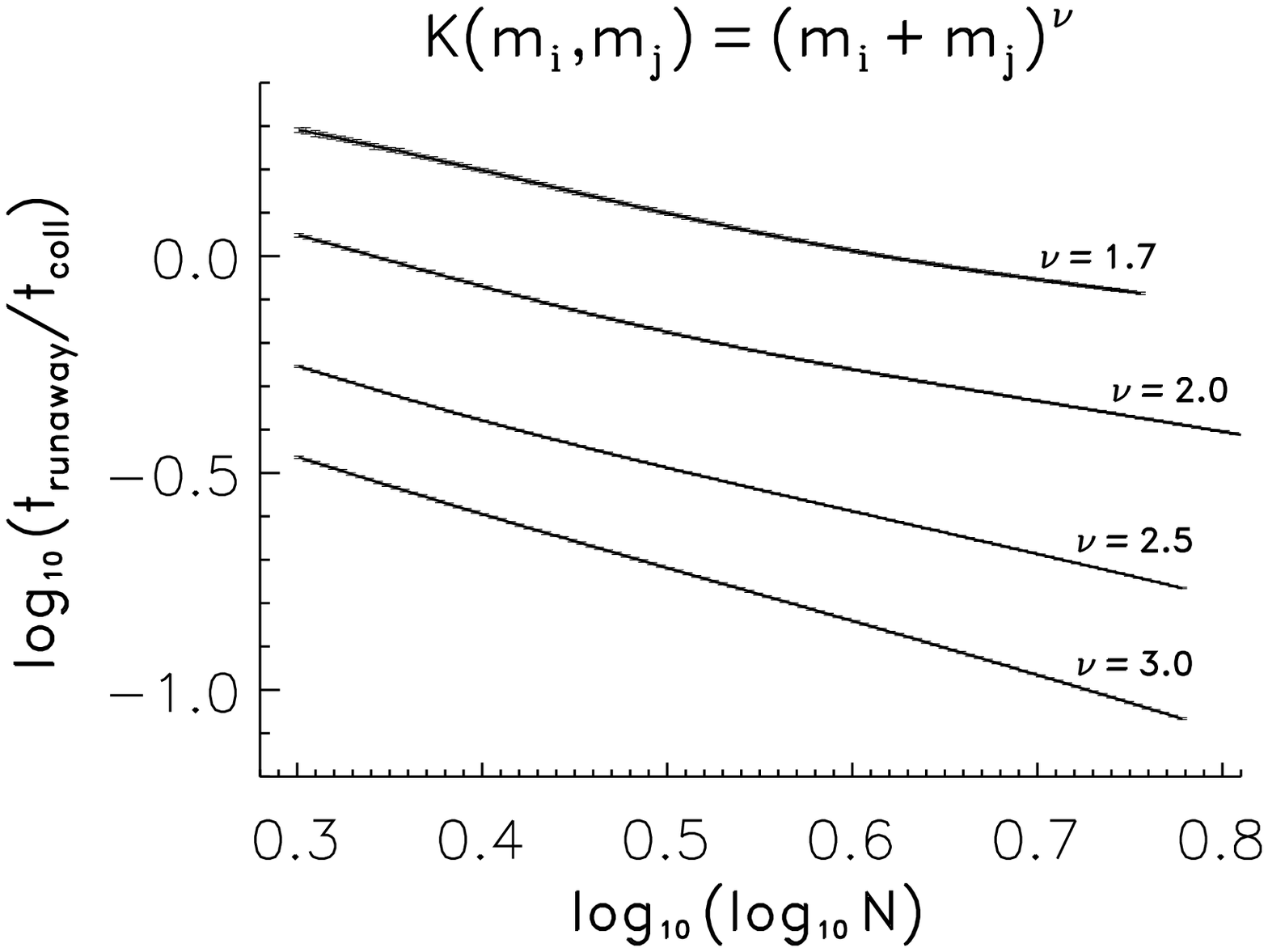}
\caption{
Average normalized time at which runaway begins versus initial
particle number $N$ for different coalescence exponents
$\nu$.  Error bars enclose $50\%$ of the simulations [$30-10^5$
simulations at each $(N,\nu)$].  }
\label{RUNAWAY_TIME}
\end{figure}

Figure~\ref{RUNAWAY_PARAMETERS} shows the evolution of various
averaged diagnostics for the representative case $\nu=2,~N=10^7$ and
for both additive and multiplicative kernels.  Instead of elapsed
time, we use the logarithm of the number of mergings, $N_{\rm merg}$, 
as the independent variable.  Fig.~\ref{RUNAWAY_PARAMETERS}(a) shows 
the (average) relationship between these coordinates.  Note that we 
have reintroduced $\tau$ [cf. eq.~(\ref{DIMENSIONAL_TIME})], which is 
time normalized to the collision time per particle in the initial state
(\ref{INITIAL_CONDITIONS}).  We see that there are two evolutionary
phases. 
In the first phase, the total coalescence rate is dominated by
mergings among unit-mass particles [Fig.~\ref{RUNAWAY_PARAMETERS}(d)].
The mass spectrum, \emph{i.e.} the distribution of average occupation
number $\overline{n}$ with particle mass $m$, is continuous, as shown
by the near-equality of the largest and second-largest
mass [Fig.~\ref{RUNAWAY_PARAMETERS}(b)], and extends to steadily higher
$m$ [Figs.~\ref{RUNAWAY_PARAMETERS}(e,c)].  The total
coalescence rate is approximately equal to its initial value
$N/2t_{\rm coll}$.  This is the \emph{statistical growth phase}.

The \emph{runaway growth phase} could be considered to begin when
there is a significant mass gap between the highest
and second-highest occupied bin [Fig.~\ref{RUNAWAY_PARAMETERS}(b)].
We find it more convenient to declare runaway when the rate
of mergings involving the highest occupied bin is $50\%$ of the total
coalescence rate [Fig.~\ref{RUNAWAY_PARAMETERS}(d)].
At this  point $m_I$, the mass per particle of the highest-mass bin,
is $\approx N^{1/\nu}$.
The runaway phase accounts for a negligible fraction of the total time 
($t$ or $\tau$) but a majority of the total mergings
 [Fig.~\ref{RUNAWAY_PARAMETERS}(a)].

For other coalescence exponents $\nu>1$, the qualitative features of
the evolution are similar to those displayed in
Fig.~\ref{RUNAWAY_PARAMETERS} at sufficiently large $N$.  As $\nu$
approaches unity, larger $N$ is required to obtain similar behavior,
and runaway begins later---\emph{i.e.}, at larger $N_{\rm merge}/N$.

At a given $N$ and $\nu$, runaway occurs later for additive than for
multiplicative kernels and involves more particles of intermediate
mass.  The latter is explained by the fact that the collision rate
between particles of very different mass is almost independent of the
lighter mass with the additive kernel.  Once the runaway phase of the
multiplicative case is well established, most of the particles are of
unit mass, and the evolution is similar to the monotrophic model
[Fig.~\ref{RUNAWAY_PARAMETERS}(e,f)].

\begin{figure}
\vspace{6.6cm}
\includegraphics{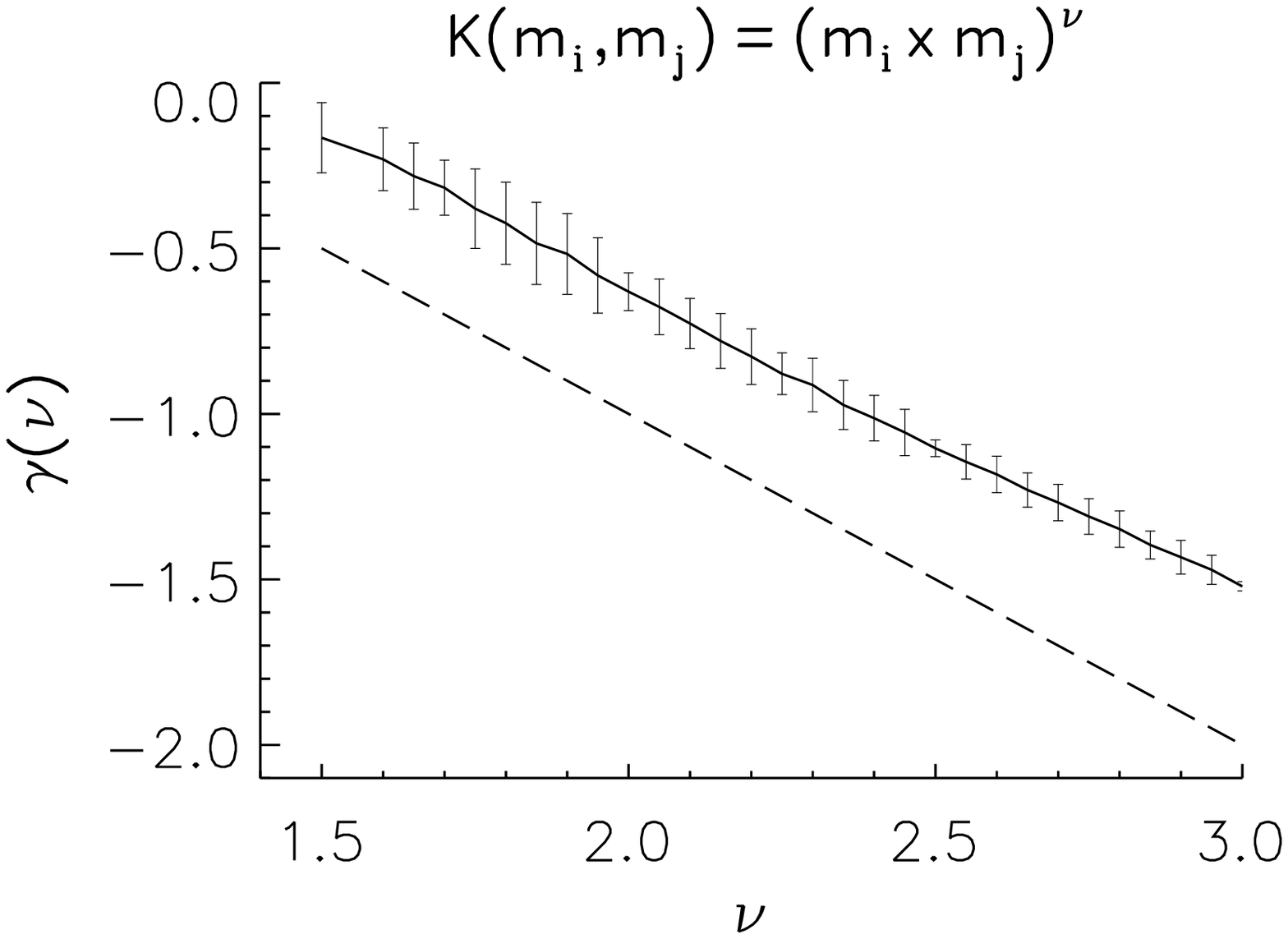}
\includegraphics{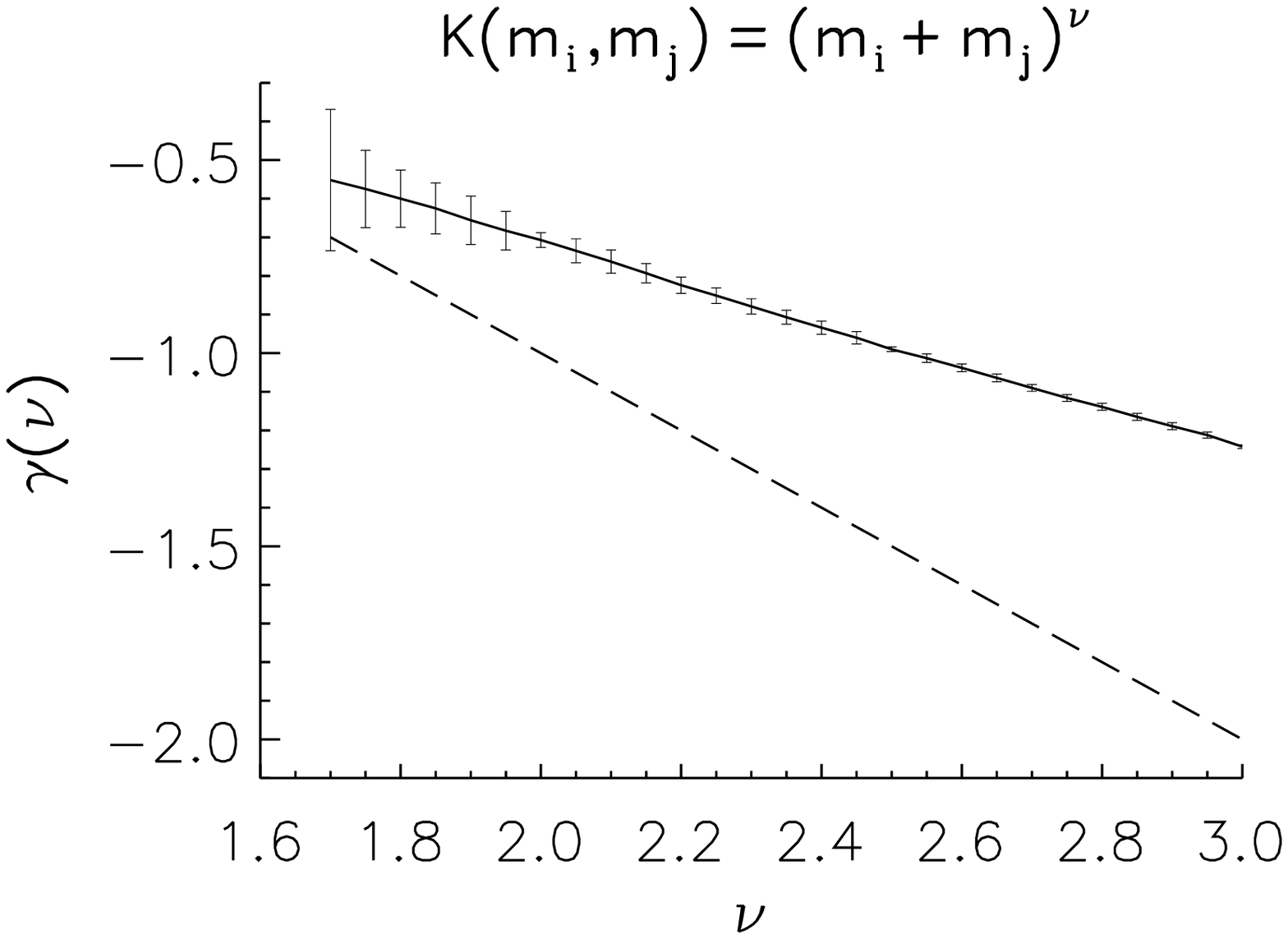}
\caption{
Exponent $\gamma$ for the relation $\tau_{\rm runaway}\propto
(\log N)^\gamma$ between runaway time and particle number, versus
coalescence exponent $\nu$.
Solid lines: best fit based on asymptotic slope seen in 
Fig.~\ref{RUNAWAY_TIME} at large $N$ (the slope and error bars are
obtained by the least-squares linear regression).
Dashed lines: prediction $\gamma=1-\nu$ of monotrophic
model [eq.~(\ref{ESTIMATED_RUNAWAY_TIME})].
}
\label{GAMMA}
\end{figure}

Figure~(\ref{RUNAWAY_TIME}) shows that the runaway time slowly decreases
with increasing $N$, even after scaling by the single-particle
collision time~(\ref{DIMENSIONAL_TIME}).  At sufficiently large $N$, it 
appears that $\tau_{\rm runaway}$ decreases as a power of $\log N$ and 
that the slope is steeper for larger $\nu$, in qualitative agreement with
the conjecture (\ref{ESTIMATED_RUNAWAY_TIME}).
The first panel of
Fig.~(\ref{GAMMA}) shows rough quantitative agreement with the conjecture
in the slope of $\gamma(\nu)$ against $\nu$ for multiplicative kernels.
There appears to be an offset in the vertical intercept, but
one knows from the exact results of \cite{Lushnikov} that
$\tau_{\rm runaway}$ is independent of $N$ as $N\to\infty$ at $\nu=1$, 
so the solid and
dashed lines should intersect at $\gamma(1)=0$.  Unfortunately, it is
extremely difficult to measure $\gamma$ near $\nu=1$ from our simulations.
The second panel of Fig.~(\ref{GAMMA}) shows that
for additive kernels, the empirical slope of $\gamma(\nu)$
is close to one half of the prediction (\ref{ESTIMATED_RUNAWAY_TIME}).


\section{DISCUSSION}\label{DISCUSSION}

Our Monte Carlo simulations indicate that runaway is instantaneous
in the limit $N\to\infty$ if the
coalescence kernel is superlinear,
in agreement with previous authors \citep{Jeon,Spouge}.
But the simulations, as well as the heuristic model of \S2, suggest a
more quantitative new result: namely, that
the limit $\tau_{\rm runaway}\to 0$ is approached only logarithmically
in $N$ for power-law kernels.

We consider briefly two astrophysical applications:
coalescence of planetesimals into planets
in a protostellar disk, and black hole formation through stellar
mergers in galactic nuclei.  In both cases, the two-body collision
rate per unit volume is $\nu_i\nu_j\langle\sigma v\rangle_{ij}$,
where $\nu_i$ is the number of particles of mass $m_i$ per unit volume,
and the rate coefficient is approximately (Lee 2000)
\be\label{RATE0}
\langle\sigma v\rangle_{ij} = \pi(R_i+R_j)^2\left[ v_{ij}
+ \frac{2G(m_i+m_j)}{(R_i+R_j)}\,\frac{1}{v_{ij}}\right].
\ee
Here $R_i$ is the radius corresponding to mass $m_i$, and $v_{ij}$ is
the root-mean-square relative velocity of the particles.
The prospects for runaway
are usually more favorable when the second (\emph{gravitational focusing})
term in the square brackets above is dominant.
Henceforth, we neglect the first term.  A power-law
mass-radius relation
\be\label{MR}
R\propto m^\alpha
\ee
describes planetesimals if  $\alpha\approx 1/3$, and stars
if $\alpha\approx 1$.
For a maxwellian velocity distribution with equipartition of kinetic energies
among mass groups, 
\bd
v_{ij}= v_1\left(\frac{i+j}{ij}\right)^{1/2},
\ed
where $v_1$ is the velocity dispersion of particles of mass $m_1$ and
$m_i=i\times m_1$. Then
\be\label{RATE1}
\langle\sigma v\rangle_{ij} \approx \frac{2\pi Gm_1 R_1}{v_1}\,
\left(i^\alpha+j^\alpha\right)(i+j)^{1/2}(ij)^{1/2}
\ee
Since this is proportional to $j^{\alpha+1}$ for $j\gg i$, runaway is
possible if $\alpha>0$.  

Mass stratification somewhat intensifies the tendency towards runaway.
A large-scale gravitational potential will concentrate the
heavier particles into a smaller volume than the lighter ones.
Thus in a disk of planetesimals with orbital angular velocity $\Omega$,
the distribution of mass group $i$ with height $z$ above the midplane is
\bd
\nu_i(z)\approx\nu_i(0)\exp\left(-z^2/2H_i^2\right),\qquad H_i\equiv
v_i/\Omega\propto i^{-1/2}.
\ed
Hence upon integrating with respect to $z$,
\be\label{STRATP}
\int\limits_{-\infty}^\infty\nu_i(z)\nu_j(z)\,dz=
\left(2\pi H_1^2\right)^{-1/2}\,\hat n_i\hat n_j\,
\left(\frac{ij}{i+j}\right)^{1/2}.
\ee
Here $\hat n_i$ is the number of particles of mass $m_i$ per unit area.
The mass-dependent factor in this expression multiplies the
rate coefficient (\ref{RATE1}) to yield an effective collision kernel
($\alpha\approx 1/3$)
\be\label{KPLANET}
K(i,j)= K(1,1)\,\times\,ij\, (i^{1/3}+j^{1/3})/2,
\ee
which has the same $j^{4/3}$ scaling for $j\gg i$ as the rate
coefficient (\ref{RATE1}) but a stronger dependence
on the mass of the lighter particle: $i$ versus $i^{1/2}$.
This is likely to cause more rapid consumption of
intermediate-mass objects, so we expect
evolution as if for a multiplicative
kernel [$K(i,j)\propto(ij)^{4/3}$] rather than an additive one
 [$K(i,j)\propto(i+j)^{4/3}$].

The situation in galactic nuclei is more complicated if the heavier
stars concentrate into a selfgravitating population.  Setting aside
this possibility (which would make runaway still more rapid) and
assuming a harmonic mean potential, we get in place of eq.~(\ref{STRATP})
\bd
\int\limits_0^\infty\nu_i(r)\nu_j(r)\, 4\pi r^2dr=
\frac{ n_i n_j}{2(2\pi)^{5/2}H_1^3}\,
\left(\frac{ij}{i+j}\right)^{3/2},
\ed
since the geometry is spherical rather than planar.  Here $n_i$ is the
total number of particles of type $i$.  So in this case, with $\alpha=1$,
\be\label{KSTAR}
K(i,j)= K(1,1)\,\times\,(ij)^2.
\ee
This is precisely our multiplicative $\nu=2$ kernel.

As many as $\sim 10^{11}$ asteroid-sized planetesimals are required to make
up the mass of a terrestrial planet.
The smallest coalescence exponent for which our simulations give reliable
averages for $\tau_{\rm runaway}$ is $\nu=3/2$, and the largest $N$
we have used is $10^9$.  Nevertheless, by extrapolating the curves
in Fig.~(\ref{RUNAWAY_TIME}), we estimate 
$t_{\rm runaway}\approx 0.4\,t_{\rm coll}$  at $N=10^{11}$, 
versus $t_{\rm runaway}\approx 2\,t_{\rm coll}$
at $N=10^2$.  The dependence on $\log N$ is somewhat stronger than one
would expect from the asymptotic scaling (\ref{ESTIMATED_RUNAWAY_TIME})
because of the curvature in Fig.~(\ref{RUNAWAY_TIME}).

Black holes associated with quasars in the nuclei of galaxies are thought
to have masses of order $10^{8-9}\,M_\odot$.  The prevailing wisdom
is that the black hole forms by accretion from a gas disk, but
formation by stellar coalescence has long been a popular alternative model
\citep[and references therein]{Rees}.  Referring again to
Fig.~(\ref{RUNAWAY_TIME}), but now for $\nu=2$, we find
$t_{\rm runaway}\approx .09\,t_{\rm coll}$ at $N=10^9$ versus
$t_{\rm runaway}\approx .35\,t_{\rm coll}$ at $N=10^2$, roughly in
agreement with the predicted asymptotic scaling $(\log N)^{1-\nu}$.



\goodbreak
\begin{appendix}
\section{Asymptotics of the monotrophic runaway probability}

To derive eqs.~(\ref{PHINU_APPROX}),
one must approximate the inverse Laplace transform
\be\label{INLAP}
\phi_\nu(\tau) = \frac{1}{2\pi i}\int\limits_{C-i\infty}^{C+i\infty} 
\tilde\phi_\nu(z) e^{z\tau}\,d\tau ,
\ee
with $\tilde\phi_\nu(z)$ represented by
the infinite product (\ref{PHIT_SOL}).
The real constant $C$ must be positive but is otherwise arbitrary,
since the integrand is analytic for $\mbox{Re(z)}>0$.
Clearly, $\tilde\phi_\nu(z)$
decreases rapidly as $\mbox{Re(z)}>0$ increases, while
$e^{z\tau}$ has the opposite trend.  A good choice for $C$ is the
point $C_0$ along the real axis where the product of these two factors
in the integrand (\ref{INLAP}) is smallest.  By analyticity, this
is actually a saddle point with respect to complex $z$,
so the integrand will further decrease along the integration contour
with increasing $|\mbox{Im}(z)|$.
For $\tau\ll 1$, $C_0\gg 1$, because the exponential varies only slowly;
whereas for $\tau\gg 1$, $C_0$ lies close to the 
pole of $\tilde\phi_\nu(z)$ at $z=0$.

Thus for small $\tau$, we are concerned with $\mbox{Re(z)}\gg 1$.
We rewrite eq.~(\ref{PHIT_SOL}) as
\be\label{LIMSUM}
\ln\left[z\tilde\phi(z)\right] = -\sum\limits_{k=1}^\infty\ln\left(1+
\frac{z}{k^\nu}\right)~=-\lim_{n\to\infty}\left[
\sum\limits_{k=1}^n\ln\left(1+\frac{k^\nu}{z}\right)
~+~\ln\,\frac{z^n}{(n!)^\nu}\right]
\ee
The latter sum is dominated by its upper limit and can
be approximated by the leading terms of the Euler-Maclaurin
formula \citep{BO78}:
\begin{eqnarray*}
\sum\limits_{k=1}^n\ln\left(1+\frac{k^\nu}{z}\right)&\approx&
\int\limits_{0}^n \ln\left(1+\frac{x^\nu}{z}\right)\,dx
~+~\frac{1}{2}\ln\left(1+\frac{n^\nu}{z}\right)\\
&=&\left(n+\frac{1}{2}\right)\ln\left(1+\frac{n^\nu}{z}\right) -n\nu
~+\nu z\int\limits_{0}^n \frac{dx}{z+x^\nu},
\end{eqnarray*}
where integration by parts has been used.
As $n\to\infty$, the latter integral becomes
\bd
\nu z\int\limits_{0}^\infty \frac{dx}{z+x^\nu} = \nu z^{1/\nu} \,
\Gamma\left(\frac{1}{\nu}\right)\Gamma\left(1-\frac{1}{\nu}\right)=
z^{1/\nu}\,\frac{\pi}{\sin(\pi/\nu)}.
\ed
After this is substituted
into eq.~(\ref{LIMSUM}) and Stirling's approximation is used for $n!$,
the terms growing with $n$ cancel, and upon taking $n\to\infty$ one has
\be\label{PHIT_APPROX}
\ln\tilde\phi(z)\approx -\frac{\pi z^{1/\nu}}{\sin(\pi/\nu)} 
-\frac{1}{2}\ln z+\frac{\nu}{2}\ln(2\pi) ~+O(z^{-1}).
\ee
Using this, we estimate
\bd
C_0 = \left[\frac{\pi}{\tau\nu\sin(\pi/\nu)}\right]^{\nu/(\nu-1)}
\ed
and then perform a straightforward steepest-descent contour integration,
\emph{i.e.} we expand $\ln\tilde\phi(z)+z\tau$ as a Taylor
series to second order in $z$ about $z=C_0$ and use the
resulting Gaussian to approximate the integrand (\ref{INLAP}).
This gives the upper line in eq.~(\ref{PHINU_APPROX}).

For $\tau\gg 1$, we evaluate (\ref{INLAP}) from the residues
of the poles at $z=0$ and $z=-1$; the remaining residues are
strongly suppressed by the exponential factor.  This yields
the lower line in eq.~(\ref{PHINU_APPROX}).
\end{appendix}

\end{document}